\documentclass[
reprint,
amsmath,amssymb,
aps,
prl,
superscriptaddress,
preprintnumbers,
floatfix,
]{revtex4-2}

\usepackage{xcolor}

\usepackage{graphicx}
\usepackage{dcolumn}
\usepackage{bm}

\begin{document}


\title{Translational and rotational irreversibility in cyclically sheared colloidal suspensions}
\thanks{A special thanks to Esmaeel Moghimi for general guidance; Ruth Crothers for synthesis guidance; Xinran Zhang and Jasper Nijdam for technical support; and Leon Der for work done on the experimental apparatus. We acknowledge the support of the National Science Foundation Grant No. DMR-2226485.}%

\author{Austin H. Walker}

\author{Jeffrey S. Urbach}%

\author{Daniel L. Blair}%

 \email{dlb76@georgetown.edu }
\affiliation{%
Department of Physics and Institute for Soft Matter Synthesis and Metrology, Georgetown University, Washington, D.C. 20057, USA
}%

\date{\today}

\begin{abstract}

We investigate the translational and rotational dynamics of cyclically sheared suspensions over a broad range of particle concentrations. The transition to an irreversible state is studied using a custom confocal-rheometer and colloidal particles synthesized with chemically identical smooth and lobed surfaces. The particle trajectories and rotations are tracked in three dimensions to reveal enhanced effective diffusivities that exhibit flow induced anisotropy that depends on particle concentration and particle morphology. We show that within the irreversible state, the shear induced anisotropy is determined by interparticle friction and interlocking mechanisms that promote mixing and enforce rotational isotropy.
 
\end{abstract}

\maketitle

 Repulsive colloidal suspensions are known to display a rich landscape of phenomena that are highly concentration dependent and linked to the physiochemical interactions and microstructure of their components. As the volume fraction $\phi$ increases, sheared repulsive suspensions undergo a transition from a state where particle interactions are dominated by hydrodynamics to a state dominated by friction \cite{Cates2014,Cohen2015,Hudson2016}. Frictional interactions are linked to several important phenomena, including shear thickening \cite{Denn2013, Morris2020, Nicos2022}, shear jamming \cite{Jaeger2020, Jaeger2023}, and flow instabilities \cite{Morris2019}. The addition of asperities (lobes) to particle surfaces can be used to enhance the frictional interactions between particles \cite{Isa2018} and to produce another type of physical interaction, the interlocking between lobes. Adding asperities to particle surfaces enhances shear thickening \cite{Isa2018} and reduces the force required to generate particle contacts near shear jamming \cite{Solomon2017}. The impact of asperities on the bulk rheological response motivates us to show how asperities effect particle dynamics and development of structure on the mesoscopic scale.

Translational and rotational dynamics are linked to particle contacts, local structure, and frictional interactions in dry-granular \cite{Wang2018} and attractive colloidal systems \cite{Dullens2021, Dullens2022}. We extend this methodology of linking dynamics to the interaction mechanisms between particles for repulsive smooth and lobed colloidal suspensions. To examine the dynamics, we leverage the time irreversibility of particle motion under oscillatory shear. At vanishing Reynolds number, particle trajectories under shear are reversible at small imposed strain \cite{Bossis1988} and become irreversible at large strain when particle interactions are sufficient to push particles off their streamline motion, enhancing particle diffusivity \cite{Leshansky2005}. Previous experiments show that changing particle roughness lowers the onset of translational irreversibility \cite{Bloen2016, Butler2023}. The lower onset for rough particles motivates us to measure the effect of particle morphology on each degree of freedom for translational motion and to map the rotational irreversibility transition to differentiate the impact of morphology on translational and rotational dynamics.

In this letter, we measure how cyclic shear modifies the diffusive dynamics of colloidal suspensions composed of smooth and lobed particles. We observe flow induced anisotropy in the degrees of freedom for translational and rotational motion that depend on $\phi$ and particle shape. Furthermore, lobed particles in contact, enhance mixing compared to smooth particles and maintain isotropic rotational degrees of freedom in spite of the applied asymmetric flow.

\begin{figure}
    \centering
\includegraphics[width=241pt, trim=0 3 0 0, clip]{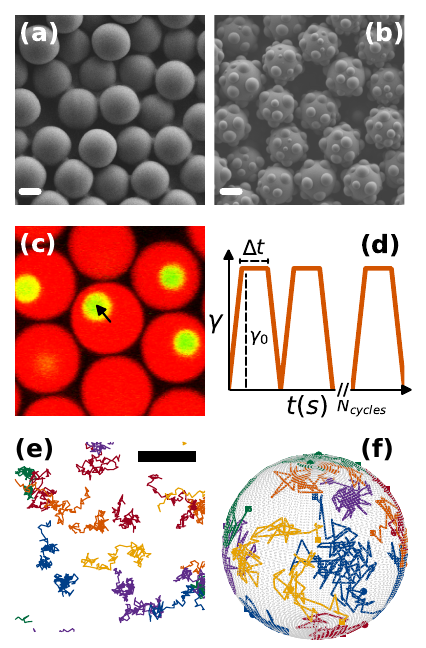} 
\caption{SEM images of smooth (a) and lobed (b) OCULI particles (scale bar 2 $\mu$m). Confocal microscopy image of OCULI particle with overlay of 2D projection of 3D orientation vector (c). The waveform of the cyclic shear applied is shown in (d). A subset of 2D translational trajectories in the flow $\hat{x}$ and vorticity $\hat{y}$ directions (e) (scale bar 3 $ \mu $m), and 3D rotational trajectories (f) for smooth particles at $\phi=42\%$, $\gamma_{0} = 0.00$.}
    \label{fig:experimental_detail}
\end{figure}

We synthesize particles with an Off-center Core Under Laser Illumination (OCULI) \cite{Dullens2019}. Each particle is composed of polymerized 3-(Trimethoxysilyl)propyl methacrylate (TPM), consisting of an inner particle (pupil) that is offset from the center of the encompassing particle (eye). The construction results in a uniform density and non-uniform fluorescence, allowing for the unbiased tracking of rotational motion. OCULI particles have been characterized and used to explore adhesion and frictional effects on dynamics in colloidal gels and colloidal glasses \cite{Dullens2021, Dullens2022}. SEM images of the smooth and lobed OCULI particles are shown in Figure \ref{fig:experimental_detail}(a,b), respectively. The OCULI synthesis protocol and characterization are detailed in the supplemental materials. The suspension is made by dispersing OCULI particles in 2,2 thiodiethanol (TDE), which is an approximate refractive index match and small density mismatch with a viscosity of  $69.6$ cp. The particles have a slight negative charge when dispersed in TDE. 

Using a confocal-rheometer \cite{Blair2013}, we apply cyclic shear defined in Figure \ref{fig:experimental_detail}(d) as a truncated triangle wave; the waveform is composed of a steady shear rate $\dot{\gamma}^{+}$ followed by locking the rheometer tool $(\dot{\gamma}=0)$ for time $\Delta t$ and finally imposing a reverse shear rate $\dot{\gamma}^{-}$. The range of diffusion times (at rest) measured for translational $\tau_{t}$ and rotational $\tau_{r}$ motion, for various sized particles and volume fractions,  are  $\tau_{t} \approx 89-714s$ and $\tau_{r} \approx 41-830s$ respectively. We fix the period of oscillation to $8\rm{s}$, much smaller than the Brownian timescale. Strain amplitudes $\gamma_{0}$ range from $0.750 - 5.00$ with Peclet numbers between $10^{2} -10^{3}$ to ensure advection dominates over thermal diffusion. Experiments were performed in the creeping-flow Stokes regime with Reynolds numbers ranging from $10^{-8}- 10^{-7}$. 

We image the sample once per cycle at the truncated peak [Fig. \ref{fig:experimental_detail}(d)], allowing us to acquire a volume that is at least $5$ particle diameters in the gradient (z) direction and $145\mu m \times 145\mu m$ in flow-vorticity $(x,y)$ plane. Before each acquisition, we apply shear to homogenize the system and ensure a reproducible initial state; we repeat the premixing followed by acquisition until three independent measurements are made.

Individual particle dynamics are determined using a suite of image analysis techniques \cite{Grier1996} resulting in time resolved three-dimensional centroids of both the eye and pupil for each particle over all times. The arrow connecting the center of the eye to the center of the pupil is a well defined vector used to identify the three-dimensional orientation of each particle in time [see Fig.\-\ref{fig:experimental_detail}(c)]. A subset of trajectories for translational motion in the flow-vorticity direction [the $(x,y)-$plane] and rotational motion centered on a unit sphere are shown in Figs.\ \ref{fig:experimental_detail}(e,f). The $(x,y,z)$ components of the $\phi$ dependent translational diffusivity $D_{T}(\phi)$ are obtained by performing least squares linear fits of the mean squared translational displacement $\langle\Delta \vec{r}^{\ 2}\rangle = \langle[\vec{r}(t + \Delta t) - \vec{r}(t)]\rangle$. Similar analysis is performed on the orientational unit vectors resulting in the three-dimensional mean squared angular displacement,
\begin{equation}
\langle \Delta\vec{\varphi}^{ 2}\rangle = \langle[\vec{\varphi}(t + \Delta t) - \vec{\varphi}(t)]^{2}  \rangle,
\end{equation}
where $\vec{\varphi}(t)$ is the orientational displacement vector and its components approximate the cumulative rotational displacement about the Cartesian axes defined above \cite{Weeks2011}. The Stokes-Einstein-Debye relation is used to extract the rotational diffusivity for all particle concentrations ${D}_{R}(\phi)$; in three-dimensions it is defined as $\langle\Delta\vec{\varphi}^{2}\rangle = 4{D}_{R}\Delta t$. Normalizing the diffusion constants extracted from the MSD and MSAD by their respective zero shear values ${D}_{(T,R)}(\dot{\gamma} = 0)$, provides the dimensionless effective translational and rotational diffusivities, ${\bf D}^3_T$ and ${\bf D}^3_R$, respectively. 

\begin{figure}
    \centering
\includegraphics[width=0.45\textwidth]{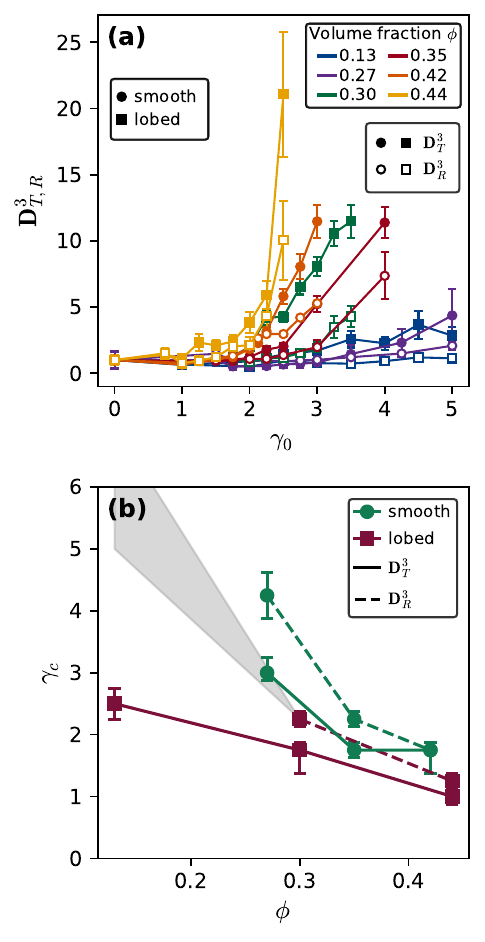} 
\caption{(a) is the 3D normalized effective diffusion coefficient for smooth (circle) and lobed (square) particles for translational and rotational motion. (b) is the critical strain value as a function of the volume fraction for smooth and lobed OCULI particles for translational and rotational motion. The shaded region highlights expected, but unmeasured transition.}
    \label{fig:3d_diffusion}
\end{figure}

Cyclic shear enhances the effective diffusivity of colloidal dispersions and athermal suspensions over a broad range of particle concentrations \cite{Leshansky2005,Pine2008}.  The magnitude of $\gamma_c$ is set by the coupling of steric colloidal interactions including surface morphology to the external shear strain. We determine $\gamma_c$ by fitting a sliding window over ${\bf D}^3_{T,R}(\gamma_0)$ and use the mid-point of the first fit line with a slope above a predetermined threshold (see Fig. S7). At strain amplitudes below a $\phi-$dependent critical value $\gamma_c$, the particles do not interact appreciably, resulting in normalized diffusivity values ${\bf D}^3(\gamma_0 < \gamma_c)_{T,R} \sim 1$ [see Fig.~\ref{fig:3d_diffusion}(a)].  It should be noted that for all of our experiments, the normalized diffusivity remains largely unaffected by strain below $\gamma_0 = 1$ even at the highest $\phi$ (see Fig.~\ref{fig:3d_diffusion}b). The external shear strain results in a deviation from reversible states where ${\bf D}^3_{T,R}\sim1$ to interaction driven kinetics that leads to a decrease in $\gamma_c$ with increasing $\phi$ (see Fig.~\ref{fig:3d_diffusion}b). 

Our protocol for particle synthesis guarantees that each experimental realization has identical short ranged physicochemical interactions (solvation, charge, and friction), leaving only concentration and surface morphology as physical variables to compare. To disentangle how steric interactions modify shear enhanced diffusion, we first focus on smooth particles at $\phi=\{0.27, 0.35, 0.42\}$. At the highest concentration ($\phi=0.42$), the critical strain for ${\bf D}^3_T$ and ${\bf D}^3_R$ are indistinguishable (Fig.~\ref{fig:3d_diffusion}b). However, as $\phi$ is reduced, the values of $\gamma_c$ separate, indicating that an increase in $\gamma_0$ is required to enhance rotational fluctuations (Fig.~\ref{fig:3d_diffusion}b). 

The addition of asperities to colloidal particles leads to dramatically altered rheological and flow behavior for concentrated suspensions \cite{Isa2013, Isa2018, Jaeger2020,10.1122/8.0000424}. Our synthesis allows us to generate randomly distributed lobes that, in the presence of shear, can mechanically interdigitate at contact, while also disrupting hydrodynamic interactions at short to mid range separations (see Fig.~\ref{fig:experimental_detail}b). We perform the same shear reversal experiments using lobed particles over a range of $\phi =\{ 0.13,0.30, 0.44\}$ revealing how morphology impacts irreversibility and enhanced diffusion. We observe that the magnitude of $\gamma_c$ for ${\bf D}^3_T$ is lower than that of smooth particles over the entire range of $\phi$ (see Fig.~\ref{fig:3d_diffusion}b). Interestingly, lobed particles also exhibit a decoupling of the transition for enhanced rotational and translational diffusion at low $\phi$ that is eliminated at the highest concentration. This suggests that the observed separation of ${\bf D}^3_T$ from ${\bf D}^3_R$ is determined primarily by concentration and a transition to frictional interactions rather than the morphology of the particles.

\begin{figure*}[htpb]
    \centering
\includegraphics[width=\textwidth]{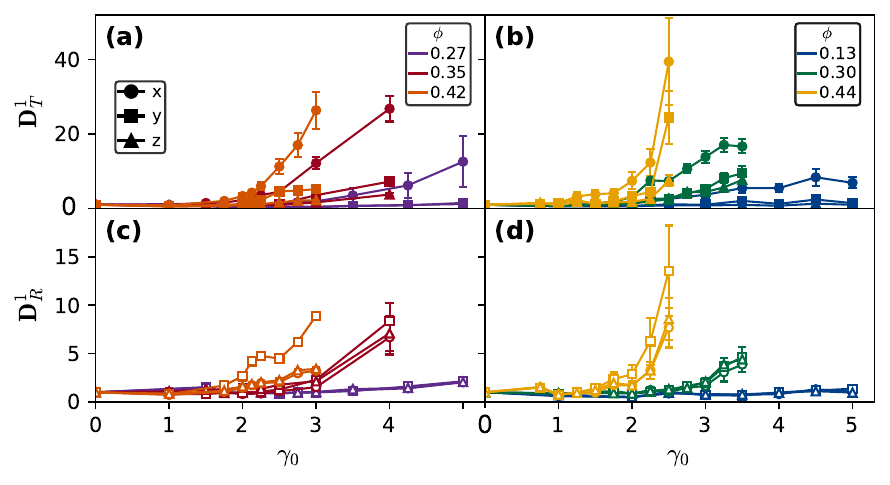} 
\caption{The one-dimensional normalized translational diffusion coefficient for smooth (a) and lobed (b) particles in the flow, vorticity, and gradient directions as a function of the applied strain amplitude is shown. The one-dimensional normalized rotational diffusion coefficient for smooth (c) and lobed (d) particles about the flow, vorticity, and gradient direction as a function of the applied strain is shown.}
    \label{one_dim}
\end{figure*}

The three-dimensional results reveal that shear strain controls the onset of irreversibility for both translational and rotational diffusivities for all surface morphologies and concentrations. However, to distinguish the impact on each degree of freedom we need to explore the one-dimensional dynamics. The normalized, one-dimensional translational diffusivities, ${\bf D}^1_{T,R}$ correspond to the rheologically orthogonal axes: $x\to$ velocity, $y\to$ vorticity, and $z\to$ gradient, whereas the rotational values correspond to the rheological axis about which the components of the orientational unit vector rotates (see Fig.~\ref{one_dim}). 

The one-dimensional data in Figure~\ref{one_dim}a reveals that above $\gamma_c$ for smooth particles, the $x-$component (velocity) of ${\bf D}^1_T$ increases at a much higher rate than the y-component (vorticity) and z-component (gradient) --  a signature of Taylor dispersion \cite{taylor_53, Leshansky2005, Chaikin2020}. A concomitant effect of the shear enhanced anisotropy is greater particle spacing along the flow direction, which we observe from the peaks of the intensity autocorrelation functions (Fig. S5). The transition to enhanced ${\bf D}^1_T$ values for lobed particles exhibits a commensurate $\gamma_{c}(\phi)$ dependence compared to smooth particles (Figure~\ref{one_dim}b). However, lobed particles exhibit a sharper transition at smaller $\gamma_{0}$ for equivalent values of $\phi$. At low $\phi$, the form of ${\bf D}^1_T$ is similar for lobed and smooth particles, whereas at the highest $\phi$, the $y-$component (vorticity) for lobed particles is greatly enhanced (Fig.~\ref{one_dim}b). The surface of smooth and lobed particles are chemically identical, indicating the severe enhancement in the vorticity direction of ${\bf D}^1_T$ for lobed particles is due to particle morphology. 
    
The new regime is linked to the lack of anisotropy in ${\bf D}^1_R$ for lobed particles (Fig.~\ref{one_dim}d). In contrast, the anisotropy in ${\bf D}^1_R$ for smooth particles is highly enhanced about the $y-$axis at $\phi=0.42$ and a similar enhancement is emerging at $\phi=0.35$ (Fig.~\ref{one_dim}c). We conjecture that  rotational anisotropy is present above $\gamma_{c}$ for smooth particles at all $\phi$, but we have not sampled high enough $\gamma_{0}$ to observe the the transition. The anisotropy is a consequence of the inherent asymmetry in the applied strain, and observationally analogous to Taylor dispersion \cite{Cohen2013, Cohen2015-2}. 

\begin{figure}
    \centering
\includegraphics[width=241pt, trim=0 0 0 0, clip]{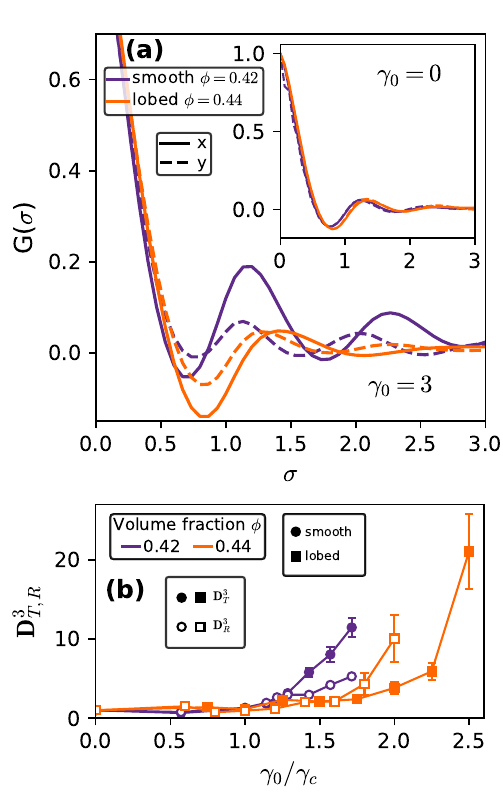} 
\caption{(a) intensity autocorrelation at 100 cycles for smooth $\phi=42\%$ (blue) and lobed $\phi=44\%$ (orange) particles at strain values $\gamma_{0} = 0.00$ $s^{-1}$ (inset) and $\gamma_{0}=3.00$ $s^{-1}$ (main) in the flow (solid) and vorticity (dashed) directions. The intensity autocorrelation in the gradient direction as been omitted for clarity. (b) the normalized 3D effective diffusion for translational and rotational motion rescaled by the critical strain amplitude.}
    \label{GR_D3}
\end{figure}

Interestingly, the enhanced rotational diffusivity, ${\bf D}^1_R$, for lobed particles remains isotropic for all $\phi$ and $\gamma_0$. To investigate if the structure reflects the differences observed in rotational isotropy, we plot the intensity autocorrelation function for smooth $(\phi=0.42)$ and lobed $(\phi=0.44)$ particles (see Fig.~\ref{GR_D3}a). As a reference, the intensity autocorrelation function for $\gamma_{0}=0$ is shown in Figure \ref{GR_D3}(inset). We observe that the peaks in the intensity autocorrelation function are greater for the smooth particles than for the lobed particles. Additionally, $G(\sigma)$ for smooth particles exhibits a broad increase in the depth of modulation, indicating an enhancement of local order which can be attributed to polycrystalline domains that are often found in cyclically sheared mono-disperse colloidal suspensions \cite{Pusey1988,Pusey1998,Petekidis2008}. The lack of strong secondary peaks for the lobed particles indicates suppression of local order. Together, the suppression of local order, isotropy in the rotational dynamics, and the enhancement of translational motion in the vorticity direction indicate enhanced mixing. Both smooth and lobed particles have similar frictional interactions, thus the morphology of the lobes suppress order and enhance mixing. 

Above the irreversibility transition for smooth and lobed particles, the growth rate of ${\bf D}^{3}_{T,R}$ is $\phi$ dependent (Fig. S6). The initial growth rate of ${\bf D}^{3}_{T,R}$ for lobed particles at $\phi=0.44$ is slower than for smooth particles at $\phi=0.42$ (see Fig.~\ref{GR_D3}b). We propose that the hindrance in ${\bf D}^{3}_{T,R}$ is due to local geometric interlocking when the lobed particles are in contact. A strain greater than $\gamma_{c}$ is needed to overcome these constraints, leading to the eventual sharpening of the transition. 

The number and nature of particle interactions drive translational and rotational irreversibility in repulsive colloidal suspensions. Our experiments reveal that the effective translational and rotational diffusivity induced by oscillatory shear can differentiate the surface mediated interactions between lobed and smooth particles. In the irreversible state, we show that flow induced anisotropy in the degrees of freedom for translational and rotational motion depend on $\phi$ and particle morphology. At high $\phi$, lobed particles maintain rotational isotropy despite imposed asymmetric flow conditions and exhibit greater enhancement along the vorticity direction compared to smooth particles. Our results indicate that increased friction from the addition of lobes is insufficient as the primary mechanism for enhancement, and that the morphology and subsequent interlocking of the lobes inhibit order, leading to enhanced mixing. Moreover, we propose the slow growth of the effective diffusion for lobed particles, at high $\phi$, is superseded by the strong enhancement in mixing once geometric constraints from the lobes are overcome by the imposed shear. In sum, particle morphology can constrain or enhance mesoscopic particle dynamics and shear induced structures that underpin the bulk rheological response of colloidal suspensions. 

\bibliographystyle{apsrev4-2}
\bibliography{papers}

\begin{thebibliography}{32}%
\makeatletter
\providecommand \@ifxundefined [1]{%
 \@ifx{#1\undefined}
}%
\providecommand \@ifnum [1]{%
 \ifnum #1\expandafter \@firstoftwo
 \else \expandafter \@secondoftwo
 \fi
}%
\providecommand \@ifx [1]{%
 \ifx #1\expandafter \@firstoftwo
 \else \expandafter \@secondoftwo
 \fi
}%
\providecommand \natexlab [1]{#1}%
\providecommand \enquote  [1]{``#1''}%
\providecommand \bibnamefont  [1]{#1}%
\providecommand \bibfnamefont [1]{#1}%
\providecommand \citenamefont [1]{#1}%
\providecommand \href@noop [0]{\@secondoftwo}%
\providecommand \href [0]{\begingroup \@sanitize@url \@href}%
\providecommand \@href[1]{\@@startlink{#1}\@@href}%
\providecommand \@@href[1]{\endgroup#1\@@endlink}%
\providecommand \@sanitize@url [0]{\catcode `\\12\catcode `\$12\catcode `\&12\catcode `\#12\catcode `\^12\catcode `\_12\catcode `\%12\relax}%
\providecommand \@@startlink[1]{}%
\providecommand \@@endlink[0]{}%
\providecommand \url  [0]{\begingroup\@sanitize@url \@url }%
\providecommand \@url [1]{\endgroup\@href {#1}{\urlprefix }}%
\providecommand \urlprefix  [0]{URL }%
\providecommand \Eprint [0]{\href }%
\providecommand \doibase [0]{https://doi.org/}%
\providecommand \selectlanguage [0]{\@gobble}%
\providecommand \bibinfo  [0]{\@secondoftwo}%
\providecommand \bibfield  [0]{\@secondoftwo}%
\providecommand \translation [1]{[#1]}%
\providecommand \BibitemOpen [0]{}%
\providecommand \bibitemStop [0]{}%
\providecommand \bibitemNoStop [0]{.\EOS\space}%
\providecommand \EOS [0]{\spacefactor3000\relax}%
\providecommand \BibitemShut  [1]{\csname bibitem#1\endcsname}%
\let\auto@bib@innerbib\@empty
\bibitem [{\citenamefont {Wyart}\ and\ \citenamefont {Cates}(2014)}]{Cates2014}%
  \BibitemOpen
  \bibfield  {author} {\bibinfo {author} {\bibfnamefont {M.}~\bibnamefont {Wyart}}\ and\ \bibinfo {author} {\bibfnamefont {M.~E.}\ \bibnamefont {Cates}},\ }\href {https://doi.org/10.1103/PhysRevLett.112.098302} {\bibfield  {journal} {\bibinfo  {journal} {Phys. Rev. Lett.}\ }\textbf {\bibinfo {volume} {112}},\ \bibinfo {pages} {098302} (\bibinfo {year} {2014})}\BibitemShut {NoStop}%
\bibitem [{\citenamefont {Lin}\ \emph {et~al.}(2015)\citenamefont {Lin}, \citenamefont {Guy}, \citenamefont {Hermes}, \citenamefont {Ness}, \citenamefont {Sun}, \citenamefont {Poon},\ and\ \citenamefont {Cohen}}]{Cohen2015}%
  \BibitemOpen
  \bibfield  {author} {\bibinfo {author} {\bibfnamefont {N.~Y.~C.}\ \bibnamefont {Lin}}, \bibinfo {author} {\bibfnamefont {B.~M.}\ \bibnamefont {Guy}}, \bibinfo {author} {\bibfnamefont {M.}~\bibnamefont {Hermes}}, \bibinfo {author} {\bibfnamefont {C.}~\bibnamefont {Ness}}, \bibinfo {author} {\bibfnamefont {J.}~\bibnamefont {Sun}}, \bibinfo {author} {\bibfnamefont {W.~C.~K.}\ \bibnamefont {Poon}},\ and\ \bibinfo {author} {\bibfnamefont {I.}~\bibnamefont {Cohen}},\ }\href {https://doi.org/10.1103/PhysRevLett.115.228304} {\bibfield  {journal} {\bibinfo  {journal} {Phys. Rev. Lett.}\ }\textbf {\bibinfo {volume} {115}},\ \bibinfo {pages} {228304} (\bibinfo {year} {2015})}\BibitemShut {NoStop}%
\bibitem [{\citenamefont {Royer}\ \emph {et~al.}(2016)\citenamefont {Royer}, \citenamefont {Blair},\ and\ \citenamefont {Hudson}}]{Hudson2016}%
  \BibitemOpen
  \bibfield  {author} {\bibinfo {author} {\bibfnamefont {J.~R.}\ \bibnamefont {Royer}}, \bibinfo {author} {\bibfnamefont {D.~L.}\ \bibnamefont {Blair}},\ and\ \bibinfo {author} {\bibfnamefont {S.~D.}\ \bibnamefont {Hudson}},\ }\href {https://doi.org/10.1103/PhysRevLett.116.188301} {\bibfield  {journal} {\bibinfo  {journal} {Phys. Rev. Lett.}\ }\textbf {\bibinfo {volume} {116}},\ \bibinfo {pages} {188301} (\bibinfo {year} {2016})}\BibitemShut {NoStop}%
\bibitem [{\citenamefont {Seto}\ \emph {et~al.}(2013)\citenamefont {Seto}, \citenamefont {Mari}, \citenamefont {Morris},\ and\ \citenamefont {Denn}}]{Denn2013}%
  \BibitemOpen
  \bibfield  {author} {\bibinfo {author} {\bibfnamefont {R.}~\bibnamefont {Seto}}, \bibinfo {author} {\bibfnamefont {R.}~\bibnamefont {Mari}}, \bibinfo {author} {\bibfnamefont {J.~F.}\ \bibnamefont {Morris}},\ and\ \bibinfo {author} {\bibfnamefont {M.~M.}\ \bibnamefont {Denn}},\ }\href {https://doi.org/10.1103/PhysRevLett.111.218301} {\bibfield  {journal} {\bibinfo  {journal} {Phys. Rev. Lett.}\ }\textbf {\bibinfo {volume} {111}},\ \bibinfo {pages} {218301} (\bibinfo {year} {2013})}\BibitemShut {NoStop}%
\bibitem [{\citenamefont {Morris}(2020)}]{Morris2020}%
  \BibitemOpen
  \bibfield  {author} {\bibinfo {author} {\bibfnamefont {J.~F.}\ \bibnamefont {Morris}},\ }\href {https://doi.org/https://doi.org/10.1146/annurev-fluid-010816-060128} {\bibfield  {journal} {\bibinfo  {journal} {Annual Review of Fluid Mechanics}\ }\textbf {\bibinfo {volume} {52}},\ \bibinfo {pages} {121} (\bibinfo {year} {2020})},\ \bibinfo {note} {publisher: Annual Reviews Type: Journal Article}\BibitemShut {NoStop}%
\bibitem [{\citenamefont {Goyal}\ \emph {et~al.}(2022)\citenamefont {Goyal}, \citenamefont {Del~Gado}, \citenamefont {Jones},\ and\ \citenamefont {Martys}}]{Nicos2022}%
  \BibitemOpen
  \bibfield  {author} {\bibinfo {author} {\bibfnamefont {A.}~\bibnamefont {Goyal}}, \bibinfo {author} {\bibfnamefont {E.}~\bibnamefont {Del~Gado}}, \bibinfo {author} {\bibfnamefont {S.~Z.}\ \bibnamefont {Jones}},\ and\ \bibinfo {author} {\bibfnamefont {N.~S.}\ \bibnamefont {Martys}},\ }\href {https://doi.org/10.1122/8.0000453} {\bibfield  {journal} {\bibinfo  {journal} {Journal of Rheology}\ }\textbf {\bibinfo {volume} {66}},\ \bibinfo {pages} {1055} (\bibinfo {year} {2022})}\BibitemShut {NoStop}%
\bibitem [{\citenamefont {Singh}\ \emph {et~al.}(2020)\citenamefont {Singh}, \citenamefont {Ness}, \citenamefont {Seto}, \citenamefont {de~Pablo},\ and\ \citenamefont {Jaeger}}]{Jaeger2020}%
  \BibitemOpen
  \bibfield  {author} {\bibinfo {author} {\bibfnamefont {A.}~\bibnamefont {Singh}}, \bibinfo {author} {\bibfnamefont {C.}~\bibnamefont {Ness}}, \bibinfo {author} {\bibfnamefont {R.}~\bibnamefont {Seto}}, \bibinfo {author} {\bibfnamefont {J.~J.}\ \bibnamefont {de~Pablo}},\ and\ \bibinfo {author} {\bibfnamefont {H.~M.}\ \bibnamefont {Jaeger}},\ }\href {https://doi.org/10.1103/PhysRevLett.124.248005} {\bibfield  {journal} {\bibinfo  {journal} {Phys. Rev. Lett.}\ }\textbf {\bibinfo {volume} {124}},\ \bibinfo {pages} {248005} (\bibinfo {year} {2020})}\BibitemShut {NoStop}%
\bibitem [{\citenamefont {Kim}\ \emph {et~al.}(2023)\citenamefont {Kim}, \citenamefont {van~der Naald}, \citenamefont {Dolinski}, \citenamefont {Rowan},\ and\ \citenamefont {Jaeger}}]{Jaeger2023}%
  \BibitemOpen
  \bibfield  {author} {\bibinfo {author} {\bibfnamefont {H.}~\bibnamefont {Kim}}, \bibinfo {author} {\bibfnamefont {M.}~\bibnamefont {van~der Naald}}, \bibinfo {author} {\bibfnamefont {N.~D.}\ \bibnamefont {Dolinski}}, \bibinfo {author} {\bibfnamefont {S.~J.}\ \bibnamefont {Rowan}},\ and\ \bibinfo {author} {\bibfnamefont {H.~M.}\ \bibnamefont {Jaeger}},\ }\href {https://doi.org/10.1039/D3SM00479A} {\bibfield  {journal} {\bibinfo  {journal} {Soft Matter}\ }\textbf {\bibinfo {volume} {19}},\ \bibinfo {pages} {6797} (\bibinfo {year} {2023})}\BibitemShut {NoStop}%
\bibitem [{\citenamefont {Seto}\ \emph {et~al.}(2019)\citenamefont {Seto}, \citenamefont {Singh}, \citenamefont {Chakraborty}, \citenamefont {Denn},\ and\ \citenamefont {Morris}}]{Morris2019}%
  \BibitemOpen
  \bibfield  {author} {\bibinfo {author} {\bibfnamefont {R.}~\bibnamefont {Seto}}, \bibinfo {author} {\bibfnamefont {A.}~\bibnamefont {Singh}}, \bibinfo {author} {\bibfnamefont {B.}~\bibnamefont {Chakraborty}}, \bibinfo {author} {\bibfnamefont {M.~M.}\ \bibnamefont {Denn}},\ and\ \bibinfo {author} {\bibfnamefont {J.~F.}\ \bibnamefont {Morris}},\ }\href {https://doi.org/10.1007/s10035-019-0931-5} {\bibfield  {journal} {\bibinfo  {journal} {Granular Matter}\ }\textbf {\bibinfo {volume} {21}},\ \bibinfo {pages} {82} (\bibinfo {year} {2019})}\BibitemShut {NoStop}%
\bibitem [{\citenamefont {Hsu}\ \emph {et~al.}(2018)\citenamefont {Hsu}, \citenamefont {Ramakrishna}, \citenamefont {Zanini}, \citenamefont {Spencer},\ and\ \citenamefont {Isa}}]{Isa2018}%
  \BibitemOpen
  \bibfield  {author} {\bibinfo {author} {\bibfnamefont {C.-P.}\ \bibnamefont {Hsu}}, \bibinfo {author} {\bibfnamefont {S.~N.}\ \bibnamefont {Ramakrishna}}, \bibinfo {author} {\bibfnamefont {M.}~\bibnamefont {Zanini}}, \bibinfo {author} {\bibfnamefont {N.~D.}\ \bibnamefont {Spencer}},\ and\ \bibinfo {author} {\bibfnamefont {L.}~\bibnamefont {Isa}},\ }\href {https://doi.org/10.1073/pnas.1801066115} {\bibfield  {journal} {\bibinfo  {journal} {Proceedings of the National Academy of Sciences}\ }\textbf {\bibinfo {volume} {115}},\ \bibinfo {pages} {5117} (\bibinfo {year} {2018})},\ \Eprint {https://arxiv.org/abs/https://www.pnas.org/doi/pdf/10.1073/pnas.1801066115} {https://www.pnas.org/doi/pdf/10.1073/pnas.1801066115} \BibitemShut {NoStop}%
\bibitem [{\citenamefont {Hsiao}\ \emph {et~al.}(2017)\citenamefont {Hsiao}, \citenamefont {Jamali}, \citenamefont {Glynos}, \citenamefont {Green}, \citenamefont {Larson},\ and\ \citenamefont {Solomon}}]{Solomon2017}%
  \BibitemOpen
  \bibfield  {author} {\bibinfo {author} {\bibfnamefont {L.~C.}\ \bibnamefont {Hsiao}}, \bibinfo {author} {\bibfnamefont {S.}~\bibnamefont {Jamali}}, \bibinfo {author} {\bibfnamefont {E.}~\bibnamefont {Glynos}}, \bibinfo {author} {\bibfnamefont {P.~F.}\ \bibnamefont {Green}}, \bibinfo {author} {\bibfnamefont {R.~G.}\ \bibnamefont {Larson}},\ and\ \bibinfo {author} {\bibfnamefont {M.~J.}\ \bibnamefont {Solomon}},\ }\href {https://doi.org/10.1103/PhysRevLett.119.158001} {\bibfield  {journal} {\bibinfo  {journal} {Phys. Rev. Lett.}\ }\textbf {\bibinfo {volume} {119}},\ \bibinfo {pages} {158001} (\bibinfo {year} {2017})}\BibitemShut {NoStop}%
\bibitem [{\citenamefont {Kou}\ \emph {et~al.}(2018)\citenamefont {Kou}, \citenamefont {Cao}, \citenamefont {Li}, \citenamefont {Xia}, \citenamefont {Li}, \citenamefont {Dong}, \citenamefont {Zhang}, \citenamefont {Zhang}, \citenamefont {Kob},\ and\ \citenamefont {Wang}}]{Wang2018}%
  \BibitemOpen
  \bibfield  {author} {\bibinfo {author} {\bibfnamefont {B.}~\bibnamefont {Kou}}, \bibinfo {author} {\bibfnamefont {Y.}~\bibnamefont {Cao}}, \bibinfo {author} {\bibfnamefont {J.}~\bibnamefont {Li}}, \bibinfo {author} {\bibfnamefont {C.}~\bibnamefont {Xia}}, \bibinfo {author} {\bibfnamefont {Z.}~\bibnamefont {Li}}, \bibinfo {author} {\bibfnamefont {H.}~\bibnamefont {Dong}}, \bibinfo {author} {\bibfnamefont {A.}~\bibnamefont {Zhang}}, \bibinfo {author} {\bibfnamefont {J.}~\bibnamefont {Zhang}}, \bibinfo {author} {\bibfnamefont {W.}~\bibnamefont {Kob}},\ and\ \bibinfo {author} {\bibfnamefont {Y.}~\bibnamefont {Wang}},\ }\href {https://doi.org/10.1103/PhysRevLett.121.018002} {\bibfield  {journal} {\bibinfo  {journal} {Phys. Rev. Lett.}\ }\textbf {\bibinfo {volume} {121}},\ \bibinfo {pages} {018002} (\bibinfo {year} {2018})}\BibitemShut {NoStop}%
\bibitem [{\citenamefont {Yanagishima}\ \emph {et~al.}(2021)\citenamefont {Yanagishima}, \citenamefont {Liu}, \citenamefont {Tanaka},\ and\ \citenamefont {Dullens}}]{Dullens2021}%
  \BibitemOpen
  \bibfield  {author} {\bibinfo {author} {\bibfnamefont {T.}~\bibnamefont {Yanagishima}}, \bibinfo {author} {\bibfnamefont {Y.}~\bibnamefont {Liu}}, \bibinfo {author} {\bibfnamefont {H.}~\bibnamefont {Tanaka}},\ and\ \bibinfo {author} {\bibfnamefont {R.~P.~A.}\ \bibnamefont {Dullens}},\ }\href {https://doi.org/10.1103/PhysRevX.11.021056} {\bibfield  {journal} {\bibinfo  {journal} {Phys. Rev. X}\ }\textbf {\bibinfo {volume} {11}},\ \bibinfo {pages} {021056} (\bibinfo {year} {2021})}\BibitemShut {NoStop}%
\bibitem [{\citenamefont {van~der Meer}\ \emph {et~al.}(2025)\citenamefont {van~der Meer}, \citenamefont {Yanagishima},\ and\ \citenamefont {Dullens}}]{Dullens2022}%
  \BibitemOpen
  \bibfield  {author} {\bibinfo {author} {\bibfnamefont {B.}~\bibnamefont {van~der Meer}}, \bibinfo {author} {\bibfnamefont {T.}~\bibnamefont {Yanagishima}},\ and\ \bibinfo {author} {\bibfnamefont {R.~P.~A.}\ \bibnamefont {Dullens}},\ }\href {https://doi.org/10.1103/PhysRevLett.134.078202} {\bibfield  {journal} {\bibinfo  {journal} {Phys. Rev. Lett.}\ }\textbf {\bibinfo {volume} {134}},\ \bibinfo {pages} {078202} (\bibinfo {year} {2025})}\BibitemShut {NoStop}%
\bibitem [{\citenamefont {Brady}\ and\ \citenamefont {Bossis}(1988)}]{Bossis1988}%
  \BibitemOpen
  \bibfield  {author} {\bibinfo {author} {\bibfnamefont {J.~F.}\ \bibnamefont {Brady}}\ and\ \bibinfo {author} {\bibfnamefont {G.}~\bibnamefont {Bossis}},\ }\href {https://doi.org/https://doi.org/10.1146/annurev.fl.20.010188.000551} {\bibfield  {journal} {\bibinfo  {journal} {Annual Review of Fluid Mechanics}\ }\textbf {\bibinfo {volume} {20}},\ \bibinfo {pages} {111} (\bibinfo {year} {1988})}\BibitemShut {NoStop}%
\bibitem [{\citenamefont {Pine}\ \emph {et~al.}(2005)\citenamefont {Pine}, \citenamefont {Gollub}, \citenamefont {F.},\ and\ \citenamefont {Leshansky}}]{Leshansky2005}%
  \BibitemOpen
  \bibfield  {author} {\bibinfo {author} {\bibfnamefont {D.~J.}\ \bibnamefont {Pine}}, \bibinfo {author} {\bibfnamefont {J.~P.}\ \bibnamefont {Gollub}}, \bibinfo {author} {\bibfnamefont {B.~J.}\ \bibnamefont {F.}},\ and\ \bibinfo {author} {\bibfnamefont {A.~M.}\ \bibnamefont {Leshansky}},\ }\href {https://doi.org/10.1038/nature04380} {\bibfield  {journal} {\bibinfo  {journal} {Nature}\ }\textbf {\bibinfo {volume} {438}},\ \bibinfo {pages} {9097} (\bibinfo {year} {2005})}\BibitemShut {NoStop}%
\bibitem [{\citenamefont {Pham}\ \emph {et~al.}(2016)\citenamefont {Pham}, \citenamefont {Butler},\ and\ \citenamefont {Metzger}}]{Bloen2016}%
  \BibitemOpen
  \bibfield  {author} {\bibinfo {author} {\bibfnamefont {P.}~\bibnamefont {Pham}}, \bibinfo {author} {\bibfnamefont {J.~E.}\ \bibnamefont {Butler}},\ and\ \bibinfo {author} {\bibfnamefont {B.}~\bibnamefont {Metzger}},\ }\href {https://doi.org/10.1103/PhysRevFluids.1.022201} {\bibfield  {journal} {\bibinfo  {journal} {Phys. Rev. Fluids}\ }\textbf {\bibinfo {volume} {1}},\ \bibinfo {pages} {022201} (\bibinfo {year} {2016})}\BibitemShut {NoStop}%
\bibitem [{\citenamefont {Zhang}\ \emph {et~al.}(2023)\citenamefont {Zhang}, \citenamefont {Pham}, \citenamefont {Metzger}, \citenamefont {Kopelevich},\ and\ \citenamefont {Butler}}]{Butler2023}%
  \BibitemOpen
  \bibfield  {author} {\bibinfo {author} {\bibfnamefont {H.}~\bibnamefont {Zhang}}, \bibinfo {author} {\bibfnamefont {P.}~\bibnamefont {Pham}}, \bibinfo {author} {\bibfnamefont {B.}~\bibnamefont {Metzger}}, \bibinfo {author} {\bibfnamefont {D.~I.}\ \bibnamefont {Kopelevich}},\ and\ \bibinfo {author} {\bibfnamefont {J.~E.}\ \bibnamefont {Butler}},\ }\href {https://doi.org/10.1103/PhysRevFluids.8.064303} {\bibfield  {journal} {\bibinfo  {journal} {Phys. Rev. Fluids}\ }\textbf {\bibinfo {volume} {8}},\ \bibinfo {pages} {064303} (\bibinfo {year} {2023})}\BibitemShut {NoStop}%
\bibitem [{\citenamefont {Liu}\ \emph {et~al.}(2019)\citenamefont {Liu}, \citenamefont {Yanagishima}, \citenamefont {Curran}, \citenamefont {Edmond}, \citenamefont {Sacanna},\ and\ \citenamefont {Dullens}}]{Dullens2019}%
  \BibitemOpen
  \bibfield  {author} {\bibinfo {author} {\bibfnamefont {Y.}~\bibnamefont {Liu}}, \bibinfo {author} {\bibfnamefont {T.}~\bibnamefont {Yanagishima}}, \bibinfo {author} {\bibfnamefont {A.}~\bibnamefont {Curran}}, \bibinfo {author} {\bibfnamefont {K.~V.}\ \bibnamefont {Edmond}}, \bibinfo {author} {\bibfnamefont {S.}~\bibnamefont {Sacanna}},\ and\ \bibinfo {author} {\bibfnamefont {R.~P.~A.}\ \bibnamefont {Dullens}},\ }\href {https://doi.org/10.1021/acs.langmuir.9b00963} {\bibfield  {journal} {\bibinfo  {journal} {Langmuir}\ }\textbf {\bibinfo {volume} {35}},\ \bibinfo {pages} {7962} (\bibinfo {year} {2019})}\BibitemShut {NoStop}%
\bibitem [{\citenamefont {Dutta}\ \emph {et~al.}(2013)\citenamefont {Dutta}, \citenamefont {Mbi}, \citenamefont {Arevalo},\ and\ \citenamefont {Blair}}]{Blair2013}%
  \BibitemOpen
  \bibfield  {author} {\bibinfo {author} {\bibfnamefont {S.~K.}\ \bibnamefont {Dutta}}, \bibinfo {author} {\bibfnamefont {A.}~\bibnamefont {Mbi}}, \bibinfo {author} {\bibfnamefont {R.~C.}\ \bibnamefont {Arevalo}},\ and\ \bibinfo {author} {\bibfnamefont {D.~L.}\ \bibnamefont {Blair}},\ }\href {https://doi.org/10.1063/1.4810015} {\bibfield  {journal} {\bibinfo  {journal} {Review of Scientific Instruments}\ }\textbf {\bibinfo {volume} {84}},\ \bibinfo {pages} {063702} (\bibinfo {year} {2013})}\BibitemShut {NoStop}%
\bibitem [{\citenamefont {Crocker}\ and\ \citenamefont {Grier}(1996)}]{Grier1996}%
  \BibitemOpen
  \bibfield  {author} {\bibinfo {author} {\bibfnamefont {J.~C.}\ \bibnamefont {Crocker}}\ and\ \bibinfo {author} {\bibfnamefont {D.~G.}\ \bibnamefont {Grier}},\ }\href {https://doi.org/https://doi.org/10.1006/jcis.1996.0217} {\bibfield  {journal} {\bibinfo  {journal} {Journal of Colloid and Interface Science}\ }\textbf {\bibinfo {volume} {179}},\ \bibinfo {pages} {298} (\bibinfo {year} {1996})}\BibitemShut {NoStop}%
\bibitem [{\citenamefont {Hunter}\ \emph {et~al.}(2011)\citenamefont {Hunter}, \citenamefont {Edmond}, \citenamefont {Elsesser},\ and\ \citenamefont {Weeks}}]{Weeks2011}%
  \BibitemOpen
  \bibfield  {author} {\bibinfo {author} {\bibfnamefont {G.}~\bibnamefont {Hunter}}, \bibinfo {author} {\bibfnamefont {K.}~\bibnamefont {Edmond}}, \bibinfo {author} {\bibfnamefont {M.}~\bibnamefont {Elsesser}},\ and\ \bibinfo {author} {\bibfnamefont {E.}~\bibnamefont {Weeks}},\ }\href {https://doi.org/10.1364/OE.19.017189} {\bibfield  {journal} {\bibinfo  {journal} {Optics Express}\ }\textbf {\bibinfo {volume} {19}},\ \bibinfo {pages} {17189} (\bibinfo {year} {2011})}\BibitemShut {NoStop}%
\bibitem [{\citenamefont {Corté}\ \emph {et~al.}(2008)\citenamefont {Corté}, \citenamefont {Chaikin}, \citenamefont {Gollub},\ and\ \citenamefont {Pine}}]{Pine2008}%
  \BibitemOpen
  \bibfield  {author} {\bibinfo {author} {\bibfnamefont {L.}~\bibnamefont {Corté}}, \bibinfo {author} {\bibfnamefont {P.~M.}\ \bibnamefont {Chaikin}}, \bibinfo {author} {\bibfnamefont {J.~P.}\ \bibnamefont {Gollub}},\ and\ \bibinfo {author} {\bibfnamefont {D.~J.}\ \bibnamefont {Pine}},\ }\href {https://doi.org/10.1038/nphys891} {\bibfield  {journal} {\bibinfo  {journal} {Nature Physics}\ }\textbf {\bibinfo {volume} {4}},\ \bibinfo {pages} {420} (\bibinfo {year} {2008})}\BibitemShut {NoStop}%
\bibitem [{\citenamefont {Fernandez}\ \emph {et~al.}(2013)\citenamefont {Fernandez}, \citenamefont {Mani}, \citenamefont {Rinaldi}, \citenamefont {Kadau}, \citenamefont {Mosquet}, \citenamefont {Lombois-Burger}, \citenamefont {Cayer-Barrioz}, \citenamefont {Herrmann}, \citenamefont {Spencer},\ and\ \citenamefont {Isa}}]{Isa2013}%
  \BibitemOpen
  \bibfield  {author} {\bibinfo {author} {\bibfnamefont {N.}~\bibnamefont {Fernandez}}, \bibinfo {author} {\bibfnamefont {R.}~\bibnamefont {Mani}}, \bibinfo {author} {\bibfnamefont {D.}~\bibnamefont {Rinaldi}}, \bibinfo {author} {\bibfnamefont {D.}~\bibnamefont {Kadau}}, \bibinfo {author} {\bibfnamefont {M.}~\bibnamefont {Mosquet}}, \bibinfo {author} {\bibfnamefont {H.}~\bibnamefont {Lombois-Burger}}, \bibinfo {author} {\bibfnamefont {J.}~\bibnamefont {Cayer-Barrioz}}, \bibinfo {author} {\bibfnamefont {H.~J.}\ \bibnamefont {Herrmann}}, \bibinfo {author} {\bibfnamefont {N.~D.}\ \bibnamefont {Spencer}},\ and\ \bibinfo {author} {\bibfnamefont {L.}~\bibnamefont {Isa}},\ }\href {https://doi.org/10.1103/PhysRevLett.111.108301} {\bibfield  {journal} {\bibinfo  {journal} {Phys. Rev. Lett.}\ }\textbf {\bibinfo {volume} {111}},\ \bibinfo {pages} {108301} (\bibinfo {year} {2013})}\BibitemShut {NoStop}%
\bibitem [{\citenamefont {Pradeep}\ \emph {et~al.}(2022)\citenamefont {Pradeep}, \citenamefont {Wessel},\ and\ \citenamefont {Hsiao}}]{10.1122/8.0000424}%
  \BibitemOpen
  \bibfield  {author} {\bibinfo {author} {\bibfnamefont {S.}~\bibnamefont {Pradeep}}, \bibinfo {author} {\bibfnamefont {A.}~\bibnamefont {Wessel}},\ and\ \bibinfo {author} {\bibfnamefont {L.~C.}\ \bibnamefont {Hsiao}},\ }\href {https://doi.org/10.1122/8.0000424} {\bibfield  {journal} {\bibinfo  {journal} {Journal of Rheology}\ }\textbf {\bibinfo {volume} {66}},\ \bibinfo {pages} {895} (\bibinfo {year} {2022})}\BibitemShut {NoStop}%
\bibitem [{\citenamefont {Taylor}(1953)}]{taylor_53}%
  \BibitemOpen
  \bibfield  {author} {\bibinfo {author} {\bibfnamefont {G.~I.}\ \bibnamefont {Taylor}},\ }\href@noop {} {\bibfield  {journal} {\bibinfo  {journal} {Proceedings of the Royal Society of London. A. Mathematical and Physical Sciences}\ }\textbf {\bibinfo {volume} {219}},\ \bibinfo {pages} {186} (\bibinfo {year} {1953})}\BibitemShut {NoStop}%
\bibitem [{\citenamefont {Wilken}\ \emph {et~al.}(2020)\citenamefont {Wilken}, \citenamefont {Guerra}, \citenamefont {Pine},\ and\ \citenamefont {Chaikin}}]{Chaikin2020}%
  \BibitemOpen
  \bibfield  {author} {\bibinfo {author} {\bibfnamefont {S.}~\bibnamefont {Wilken}}, \bibinfo {author} {\bibfnamefont {R.~E.}\ \bibnamefont {Guerra}}, \bibinfo {author} {\bibfnamefont {D.~J.}\ \bibnamefont {Pine}},\ and\ \bibinfo {author} {\bibfnamefont {P.~M.}\ \bibnamefont {Chaikin}},\ }\href {https://doi.org/10.1103/PhysRevLett.125.148001} {\bibfield  {journal} {\bibinfo  {journal} {Phys. Rev. Lett.}\ }\textbf {\bibinfo {volume} {125}},\ \bibinfo {pages} {148001} (\bibinfo {year} {2020})}\BibitemShut {NoStop}%
\bibitem [{\citenamefont {Leahy}\ \emph {et~al.}(2013)\citenamefont {Leahy}, \citenamefont {Cheng}, \citenamefont {Ong}, \citenamefont {Liddell-Watson},\ and\ \citenamefont {Cohen}}]{Cohen2013}%
  \BibitemOpen
  \bibfield  {author} {\bibinfo {author} {\bibfnamefont {B.~D.}\ \bibnamefont {Leahy}}, \bibinfo {author} {\bibfnamefont {X.}~\bibnamefont {Cheng}}, \bibinfo {author} {\bibfnamefont {D.~C.}\ \bibnamefont {Ong}}, \bibinfo {author} {\bibfnamefont {C.}~\bibnamefont {Liddell-Watson}},\ and\ \bibinfo {author} {\bibfnamefont {I.}~\bibnamefont {Cohen}},\ }\href {https://doi.org/10.1103/PhysRevLett.110.228301} {\bibfield  {journal} {\bibinfo  {journal} {Phys. Rev. Lett.}\ }\textbf {\bibinfo {volume} {110}},\ \bibinfo {pages} {228301} (\bibinfo {year} {2013})}\BibitemShut {NoStop}%
\bibitem [{\citenamefont {Leahy}\ \emph {et~al.}(2015)\citenamefont {Leahy}, \citenamefont {Koch},\ and\ \citenamefont {Cohen}}]{Cohen2015-2}%
  \BibitemOpen
  \bibfield  {author} {\bibinfo {author} {\bibfnamefont {B.~D.}\ \bibnamefont {Leahy}}, \bibinfo {author} {\bibfnamefont {D.~L.}\ \bibnamefont {Koch}},\ and\ \bibinfo {author} {\bibfnamefont {I.}~\bibnamefont {Cohen}},\ }\href {https://doi.org/10.1017/jfm.2015.186} {\bibfield  {journal} {\bibinfo  {journal} {Journal of Fluid Mechanics}\ }\textbf {\bibinfo {volume} {772}},\ \bibinfo {pages} {42–79} (\bibinfo {year} {2015})}\BibitemShut {NoStop}%
\bibitem [{\citenamefont {Ackerson}\ and\ \citenamefont {Pusey}(1988)}]{Pusey1988}%
  \BibitemOpen
  \bibfield  {author} {\bibinfo {author} {\bibfnamefont {B.~J.}\ \bibnamefont {Ackerson}}\ and\ \bibinfo {author} {\bibfnamefont {P.~N.}\ \bibnamefont {Pusey}},\ }\href {https://doi.org/10.1103/PhysRevLett.61.1033} {\bibfield  {journal} {\bibinfo  {journal} {Phys. Rev. Lett.}\ }\textbf {\bibinfo {volume} {61}},\ \bibinfo {pages} {1033} (\bibinfo {year} {1988})}\BibitemShut {NoStop}%
\bibitem [{\citenamefont {Haw}\ \emph {et~al.}(1998)\citenamefont {Haw}, \citenamefont {Poon},\ and\ \citenamefont {Pusey}}]{Pusey1998}%
  \BibitemOpen
  \bibfield  {author} {\bibinfo {author} {\bibfnamefont {M.~D.}\ \bibnamefont {Haw}}, \bibinfo {author} {\bibfnamefont {W.~C.~K.}\ \bibnamefont {Poon}},\ and\ \bibinfo {author} {\bibfnamefont {P.~N.}\ \bibnamefont {Pusey}},\ }\href {https://doi.org/10.1103/PhysRevE.57.6859} {\bibfield  {journal} {\bibinfo  {journal} {Phys. Rev. E}\ }\textbf {\bibinfo {volume} {57}},\ \bibinfo {pages} {6859} (\bibinfo {year} {1998})}\BibitemShut {NoStop}%
\bibitem [{\citenamefont {Koumakis}\ \emph {et~al.}(2008)\citenamefont {Koumakis}, \citenamefont {Schofield},\ and\ \citenamefont {Petekidis}}]{Petekidis2008}%
  \BibitemOpen
  \bibfield  {author} {\bibinfo {author} {\bibfnamefont {N.}~\bibnamefont {Koumakis}}, \bibinfo {author} {\bibfnamefont {A.~B.}\ \bibnamefont {Schofield}},\ and\ \bibinfo {author} {\bibfnamefont {G.}~\bibnamefont {Petekidis}},\ }\href {https://doi.org/10.1039/b805171b} {\bibfield  {journal} {\bibinfo  {journal} {Soft Matter}\ }\textbf {\bibinfo {volume} {4}},\ \bibinfo {pages} {2008} (\bibinfo {year} {2008})}\BibitemShut {NoStop}%
\end{thebibliography}%
\end{document}


\section{OCULI synthesis}

Colloidal particles with an off-center core under laser illumination (OCULI) \cite{Dullens2019} and uniform density were synthesized to enable tracking of rotational motion. The off-center core (pupil) and encompassing particle (eye) consist of 3-(Trimethoxysilyl)propyl methacrylate (TPM). To construct the pupils, we first hydrolyze a monomer solution of TPM by adding $14.25ml$ MilliQ + $0.75ml$ $10mM$ HCl + $1.5mL$ TPM to a glass vial and stir for one hour or until the solution becomes clear. Second, we dilute hydrolyzed TPM (hTPM) into $0.5mM$ HCl in a $100\mu m l$ round-bottom flask and stir at $600rpm$ for $10$ minutes. The exact ratio of hTPM to $0.5mM$ HCl can be adjusted to attain the desired particle size. Under quiescent conditions, we add $25ml$ $0.028\%$ $NH_{4}OH$ as one volume, making sure not to disturb the solution. Within $45$ minutes, droplets of hTPM should nucleate and the solution should turn white. Third, we add $200\mu l$ of dye solution (BDP NHS Ester, 3-Aminopropylsilane, Chloroform) and stir for one hour. Fourth, we add Azobisisobutyronitrile (AIBN) and stir for $30$ minutes. Fifth, we transfer the solution to $50ml$ falcon tubes and bake at $80C^{\circ}$ for 3 hours, making sure to tumble every $30min$. The addition of Azobisisobutyronitrile (AIBN) followed by a baking procedure polymerizes the pupils through free radical polymerization.

To synthesize OCULI particles, we start by prewetting $2 ml$ of polymerized pupils in a solution of $50 \mu l -200 \mu l$ hTPM and $5ml$ $0.028\%$ $NH_{4}OH$ for $30min$. The optimum concentration of hTPM will have the fewest secondary particles formed. Dilute wetted pupil solution to $20ml$ of $0.028\%$ $NH_{4}OH$ in a round-bottom flask and stir. Then, drip hTPM at $100\mu l/min$ to the solution until particle sizes are $2.5-3$ times larger than the pupils. If making lobed particles, add $5ml$ $5\%wt$ Pluronic F108 and stir for $30min$; if you want smooth particles without F108, skip adding the surfactant. Add $200\mu l$ of dye solution (Cyanine 3 NHS ester, 4-Aminostyrene, Acetone) and stir for one hour. We repeat the addition of Azobisisobutyronitrile (AIBN) and the baking procedure. 

To form lobes on the surface of particles, we resuspend Pluronic F108-coated OCULI particles into $11ml$ of MilliQ. Then dilute $10ml$ of particles in $500ml$ $1mM$ $NaOH$ and $50 \mu l - 200 \mu l$ of hTPM and bake for $45min$. An ideal concentration of hTPM will result in symmetric lobes with little secondary particle formation. To polymerize, we repeat the addition of Azobisisobutyronitrile (AIBN) and the baking procedure. The residual Pluronic F108 is washed out.

OCULI particles are washed in 2,2-thiodiethanol (TDE) at least 3 times to form a suspension. The refractive index of TPM, 1.51 \cite{Kraft2017}, is a close match to the refractive index of TDE, $1.52$. The density mismatch between the OCULI particles and the solvent is approximately $7\%$. We measured the viscosity of the TDE to be $48$ $cp$ indicating some atmospheric water has been absorbed. Furthermore, we report the concentration by directly measuring the mass fraction and converting it to volume fraction using the known density of TDE and cross-linked TPM.

\section{Pre-Shear Protocol}

We apply a constant shear rate to homogenize the system between experiments. The constant shear rates used to homogenize the system varied. As the volume fraction increased, we found higher rates for longer times were needed to remove memory of previous runs. We used visual conformation to ensure that cyclic shear induced ordering was removed by the constant pre-shear. When needed we ramped down to vanishing rates followed by a pause to eliminate residual drift from pre-shear.

\section{OCULI characterization}

\begin{figure}[h!]
    \centering
     \includegraphics[width=0.75\textwidth, trim=125 240 130 250, clip]{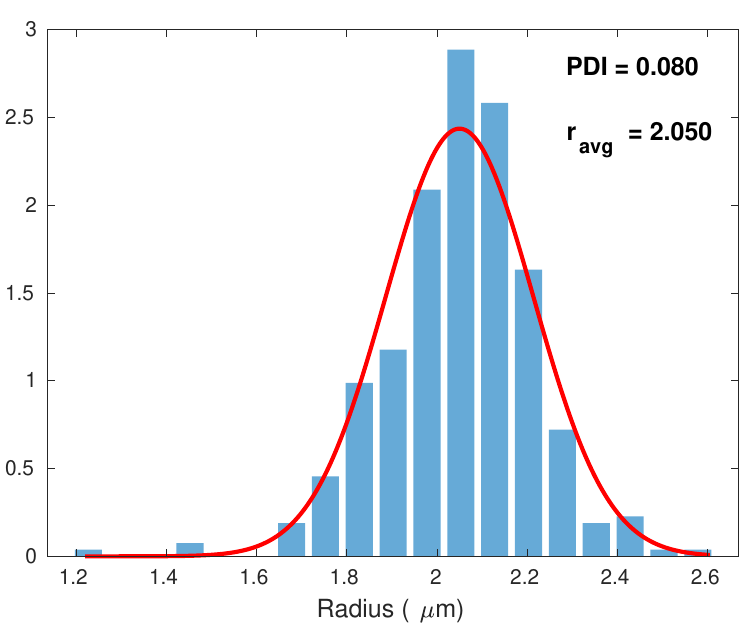} 
    \caption{Size distribution of smooth particles (batch 1).}
    \label{fig:1}

\end{figure}

\begin{figure}[h!]

    \begin{subfigure}{0.48\textwidth}
        \centering
        \includegraphics[width=\textwidth, trim=135 240 130 250, clip]{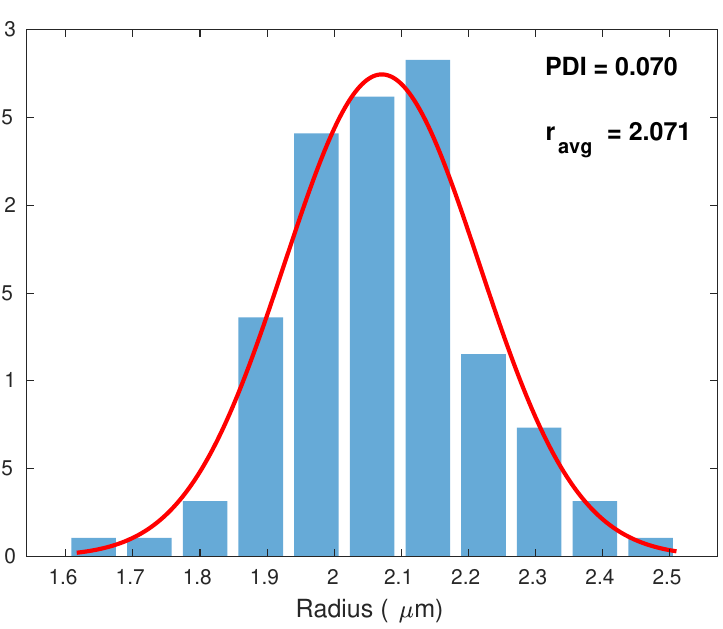} 
        \vspace{-12pt}
         \label{(a)}
         \caption{}
    \end{subfigure}
    \hfill
    \begin{subfigure}{0.48\textwidth}
        \centering
        \includegraphics[width=\textwidth, trim=135 240 130 250, clip]{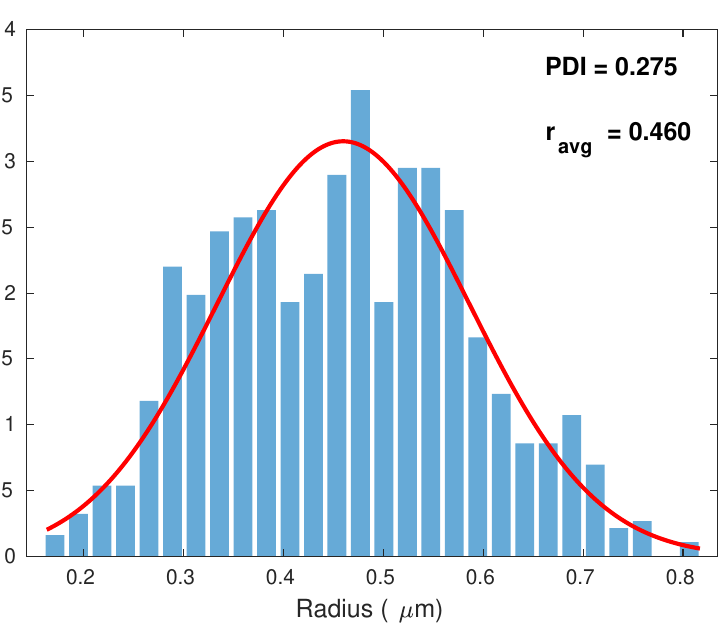} 
        \vspace{-12pt}
         \label{(b)}
         \caption{}
    \end{subfigure}
    \vspace{12pt}
    \hfill
    \begin{subfigure}{0.48\textwidth}
        \centering
        \includegraphics[width=\textwidth, trim=135 240 130 250, clip]{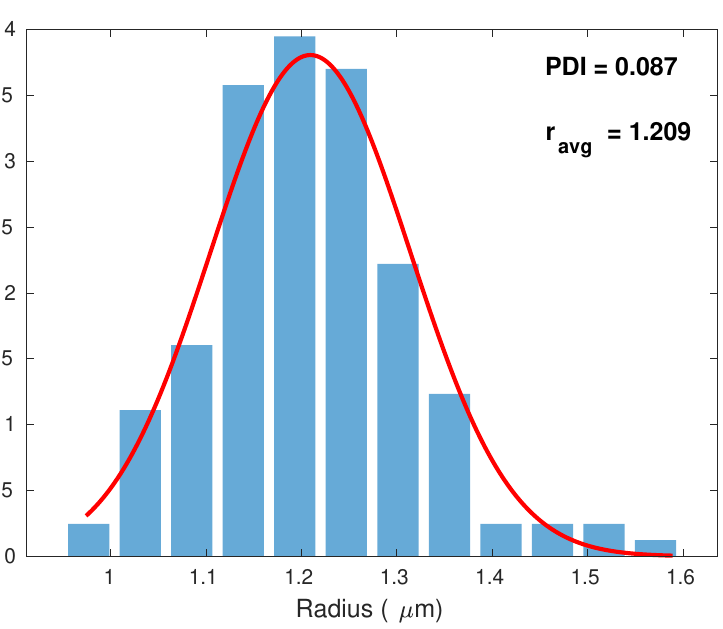} 
        \vspace{-12pt}
         \label{(c)}
         \caption{}
    \end{subfigure}
    \hfill
    \begin{subfigure}{0.48\textwidth}
        \centering
        \includegraphics[width=\textwidth, trim=135 240 130 250, clip]{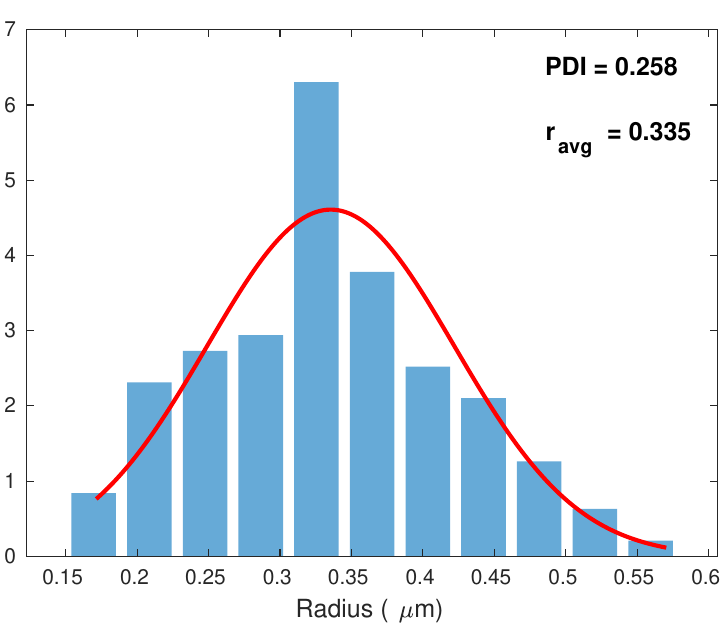} 
         \vspace{-12pt}
         \label{(d)}
          \caption{}
    \end{subfigure}
    \caption{Size distribution of lobed particles (radius, excluding contribution from lobes) and particle polydispersity index (PDI) for batch 2 (a) and batch 3 (c). Size distribution of lobes  and lobe polydispersity index (PDI) for batch 2 (b) and batch 3 (d). }
    \label{fig:2}
\end{figure}

Smooth and lobed OCULI particles are sized using SEM imaging. The size distributions for particles used are shown in figure \ref{fig:1}, and figure \ref{fig:2}. The smooth particles synthesized have a radius $R_{s}=2.05 \mu m$ and a polydispersity index of $8\%$. The two batches of lobed particles used were measured to have radii $R_{a} = 2.07\mu m$ and $R_b=1.21 \mu m $ with a polydispersity index of $7\%$ and $9\%$, respectively. The roughness of the lobed particles can be characterized as the radius of the lobes divided by the radius of the body. The two batches of lobed particles have similar roughness at  $R_{a_{lobes}}/R_{a_{body}} \approx 0.22$  and $R_{b_{lobes}}/R_{b_{body}} \approx 0.28$. Although the lobes are spread symmetrically across the surface of the particles, their high polydispersity, ranging from $26\% - 28\%$, leads to enhanced randomization of the interactions.

\clearpage

\section{Translational and rotational dynamics}

\begin{figure}[h!]
    \centering
        \includegraphics[width=241pt, trim=0 5 0 0, clip]{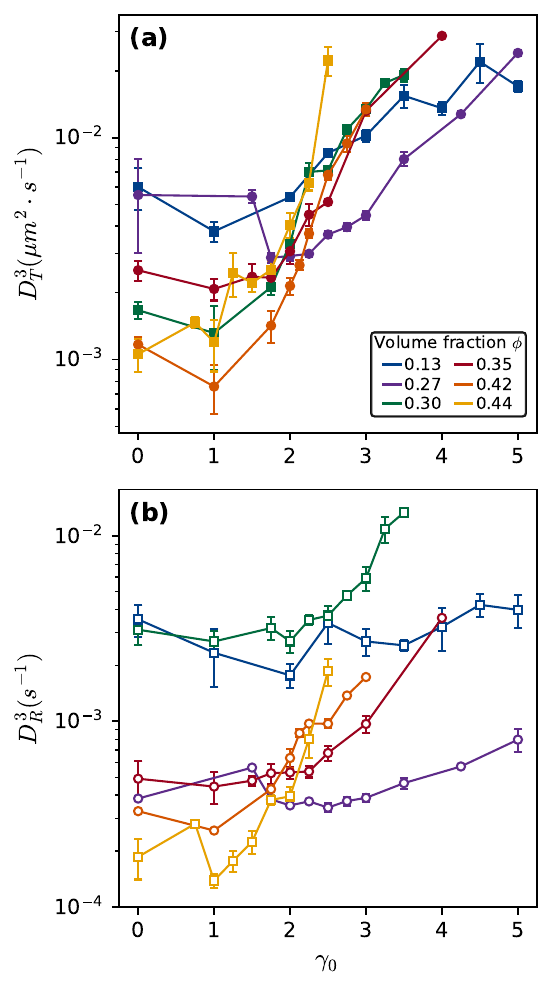} 
    \caption{The unscaled three-dimensional effective translational (a) and rotational (b) diffusion. Lobed particles (square) at $ \phi = \left\{0.13,0.30 \right\}$, exhibit faster diffusion due to smaller size. The Stokes-Einstein equation predicts an approximate match for the measured diffusion at rest. Small differences between the theoretical and measured values are likely due to the hygroscopic nature of TDE.}
    \label{fig:3}
\end{figure}

\begin{figure}[h!]
    \centering
        \includegraphics[width=241pt, trim=0 5 0 0, clip]{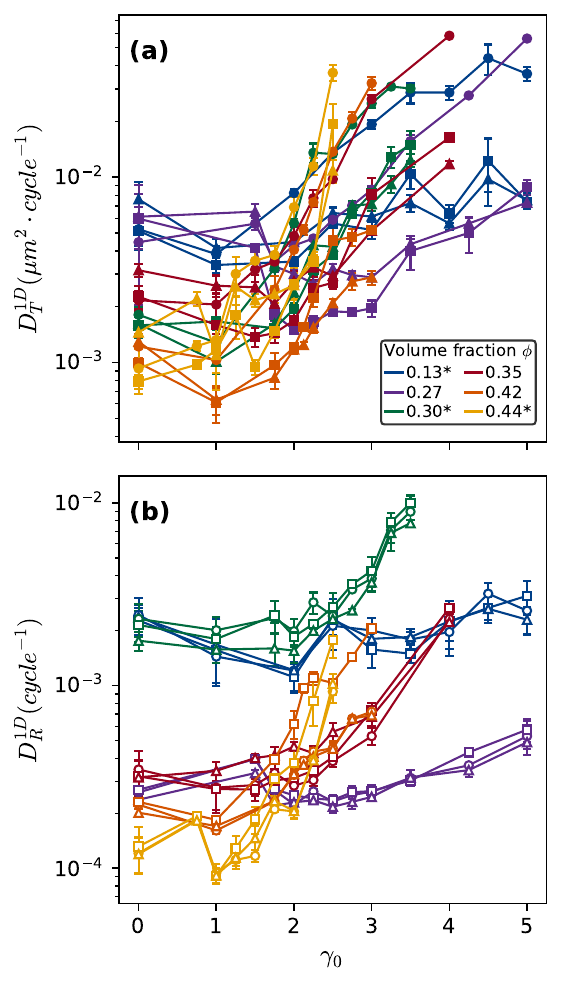} 
    \caption{The unscaled one-dimensional effective translational (a) and rotational (b) diffusion for respective rheological axes: flow (circle), vorticity (square), and gradient (triangle) .Lobed particles at $ \phi = \left\{  0.13,0.30 \right\}$, exhibit faster diffusion due to smaller size. The (*) denotes lobed particles.}
    \label{fig:4}
\end{figure}

\clearpage
\section{Structure}

\begin{figure}[h!]
    \centering
    \includegraphics[width=241pt, trim=0 5 0 0, clip]{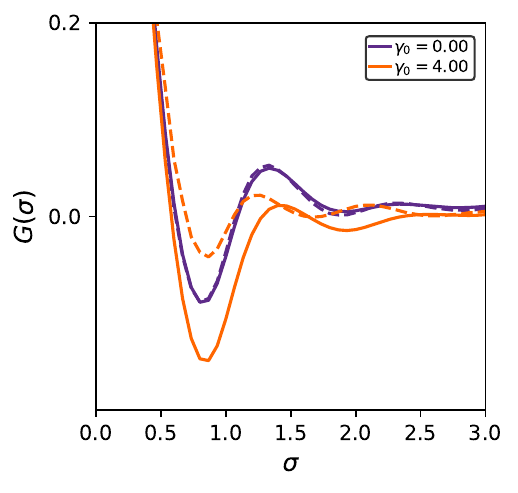} 
    \caption{Intensity autocorrelation function for smooth OCULI particles at $\phi=0.35$ in the flow (solid) and vorticity (dashed) direction at 100 cycles. Under oscillatory shear, the interparticle spacing increases in the flow direction due to Taylor dispersion.}
    \label{fig:5}

\end{figure}
    \begin{figure}[h!]
    \centering
    \includegraphics[width=241pt, trim=0 5 0 0, clip]{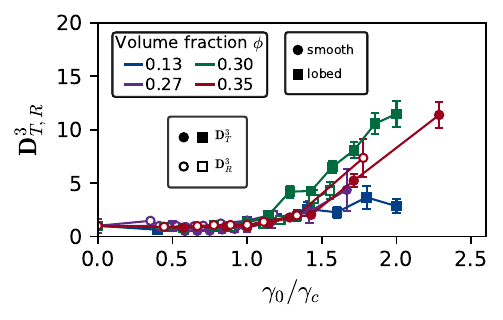} 
    \caption{The normalized 3D effective diffusion for translational and rotational motion rescaled by the critical strain amplitude.}
    \label{fig:6}

\end{figure}

\begin{figure}[h!]
    \centering
    \includegraphics[width=241pt, trim=0 5 0 0, clip]{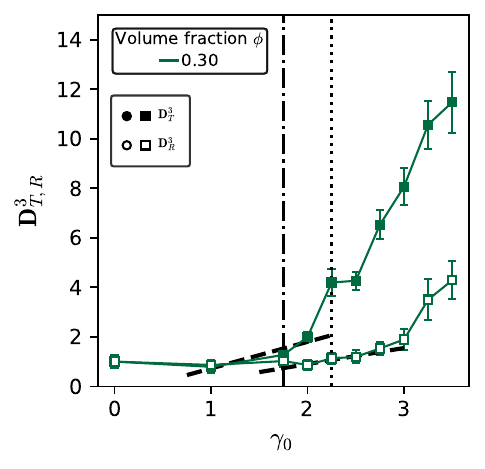} 
    \caption{The normalized 3D effective diffusion for translational and rotational motion. The dashed lines show the first sliding window linear fit where the slope exceeds 0.5. The vertical dashed-dot and and dot-dot lines denote the critical strain amplitude for translational and rotational motion respectively. All $\gamma_c$ values were calculated using same threshold for slope.}
    \label{fig:7}

\end{figure}

\begin{figure}[h!]
    \begin{subfigure}{0.45\textwidth}
        \centering
        \includegraphics[width=\textwidth, trim=105 235 125 255, clip]{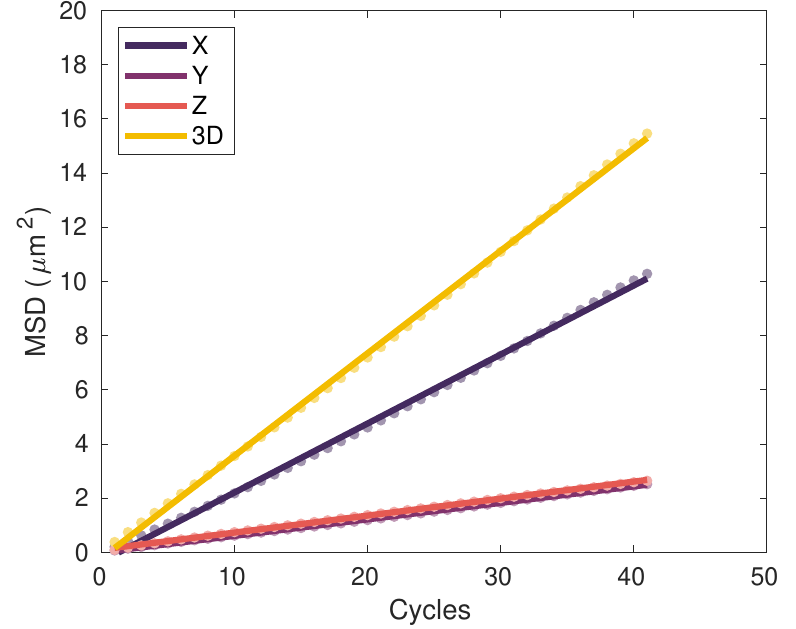} 
        \caption{}
    \end{subfigure}
        \begin{subfigure}{0.45\textwidth}
        \centering
        \includegraphics[width=\textwidth, trim=105 235 125 255, clip]{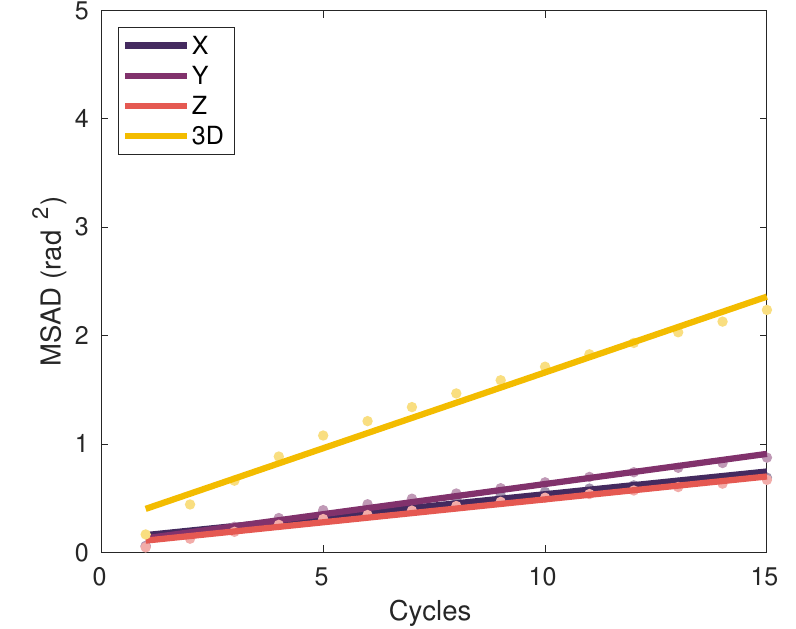} 
        \caption{}
    \end{subfigure}
    \hfill
    \begin{subfigure}{0.45\textwidth}
        \centering
        \includegraphics[width=\textwidth, trim=105 235 125 255, clip]{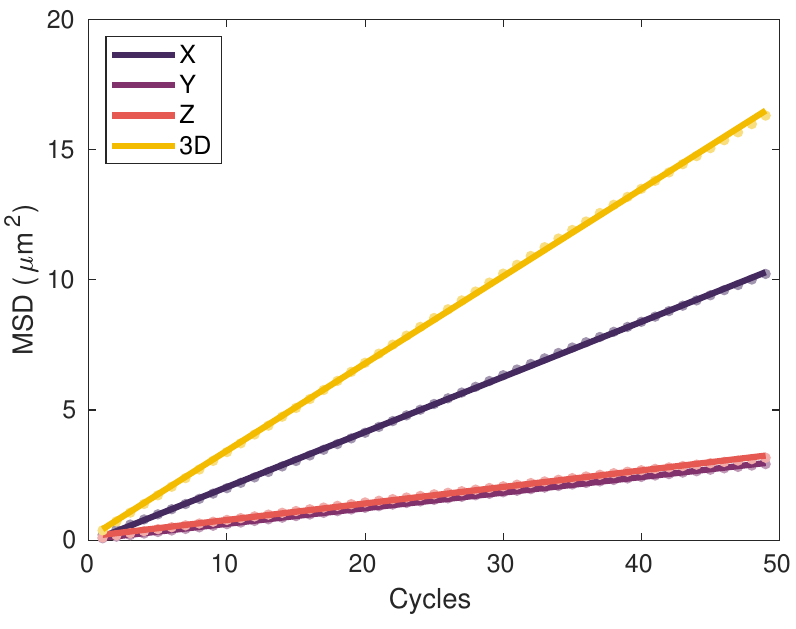} 
         \caption{}
    \end{subfigure}
       \begin{subfigure}{0.45\textwidth}
        \centering
        \includegraphics[width=\textwidth, trim=105 235 125 255, clip]{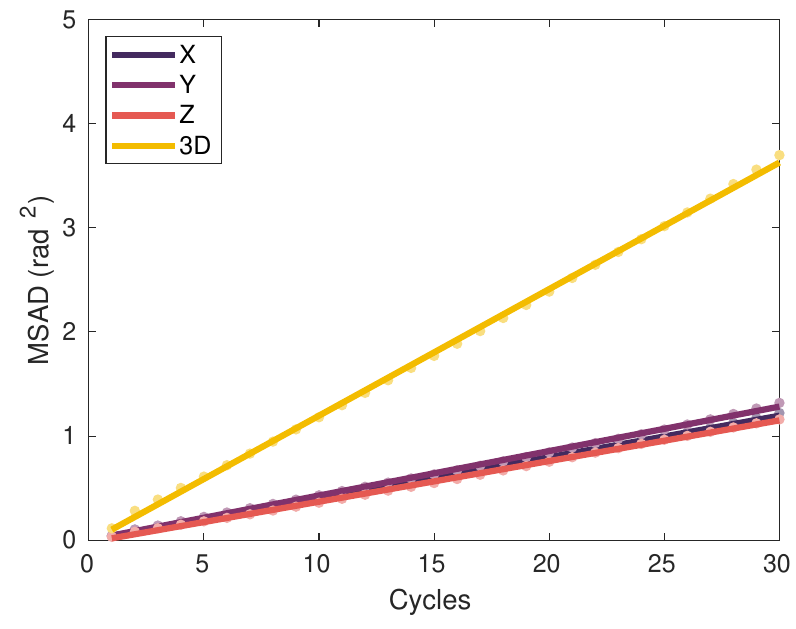}
         \caption{}

    \end{subfigure}
    \hfill
    \begin{subfigure}{0.45\textwidth}
        \centering
        \includegraphics[width=\textwidth, trim=105 235 125 255, clip]{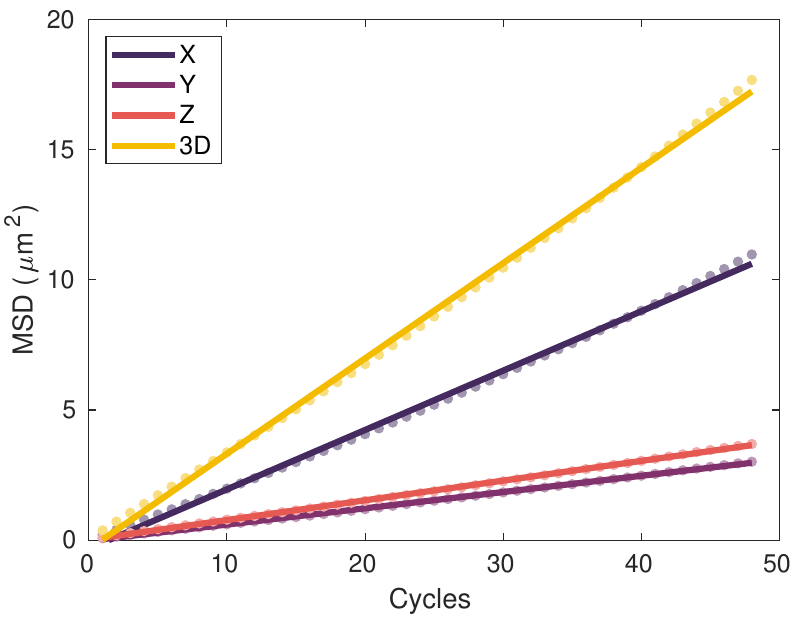} 
        \caption{}
     \end{subfigure}
        \hfill
    \begin{subfigure}{0.45\textwidth}
        \centering
        \includegraphics[width=\textwidth, trim=105 235 125 255, clip]{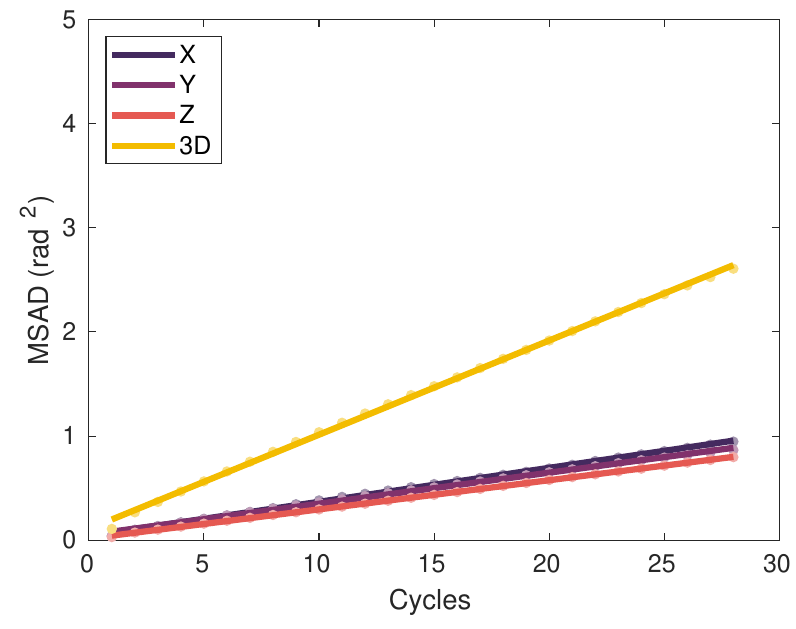} 
        \caption{}
    \end{subfigure}
    \caption{The mean squared displacement (MSD) and mean squared angular displacement (MSAD) are shown in subplots (a,c,e) and (b,d,f) for three independent imaging acquisitions for lobed particles at $\phi=0.30$ with applied strain amplitude $\gamma_0=2.5$. The lines show the fits used to calculate the ${\bf D}^3_{T,R}$, for 3D and ${\bf D}^1_{T,R}$ in 1D for the flow (x), vorticity (y), and gradient (z) directions.}
    \label{fig:8}
\end{figure}

\clearpage
\bibliographystyle{plain}
\bibliography{papers}